\documentclass[a4paper,fleqn,usenatbib]{mnras}

\usepackage{newtxtext,newtxmath}

\usepackage[T1]{fontenc}
\usepackage{ae,aecompl}

\usepackage{graphicx}	
\usepackage{amsmath}	
\usepackage{amssymb}	
\usepackage[british]{babel}
\usepackage{multicol}
\graphicspath{{/Users/brandyn/Documents/latex/images/}}

\title[Impact of baryons on lensing by clusters]{The relative impact of baryons and cluster shape on weak lensing mass estimates of galaxy clusters}

\author[B. E. Lee et al.]{
B. E. Lee,$^{1}$\thanks{E-mail: bel072000@utdallas.edu}
A. M. C. Le Brun,$^{2,3}$
M. E. Haq,$^{1}$
N. J. Deering,$^{1}$ 
L. J. King,$^{1}$
\newauthor{D. Applegate}$^{4}$
and I. G. McCarthy$^{5}$ 
\\
$^{1}$Department of Physics, The University of Texas at Dallas, 800 W Campbell Rd, Richardson, TX, 75080, USA\\
$^{2}$IRFU, CEA, Université Paris-Saclay, F-91191 Gif-sur-Yvette, France\\
$^{3}$Université Paris Diderot, AIM, Sorbonne Paris Cité, CEA, CNRS, F-91191 Gif-sur-Yvette, France\\
$^{4}$Kavli Institute for Cosmological Physics, University of Chicago, Chicago, IL, 60637, USA\\
$^{5}$Astrophysics Research Institute, Liverpool John Moores University, 146 Brownlow Hill, Liverpool L3 5RF, UK
}

\date{Accepted XXX. Received YYY; in original form ZZZ}

\pubyear{2018}

\begin{document}
\label{firstpage}
\pagerange{\pageref{firstpage}--\pageref{lastpage}}
\maketitle

\begin{abstract}
Weak gravitational lensing depends on the integrated mass along the line of sight. Baryons contribute to the mass distribution of galaxy clusters and the resulting mass estimates from lensing analysis. We use the cosmo-OWLS suite of hydrodynamic simulations to investigate the impact of baryonic processes on the bias and scatter of weak lensing mass estimates of clusters. These estimates are obtained by fitting NFW profiles to mock data using MCMC techniques. In particular, we examine the difference in estimates between dark matter-only runs and those including various prescriptions for baryonic physics. We find no significant difference in the mass bias when baryonic physics is included, though the overall mass estimates are suppressed when feedback from AGN is included. For lowest-mass systems for which a reliable mass can be obtained ($M_{200} \approx 2 \times 10^{14}$ $M_{\odot}$), we find a bias of $\approx -10$ per cent. The magnitude of the bias tends to decrease for higher mass clusters, consistent with no bias for the most massive clusters which have masses comparable to those found in the CLASH and HFF samples. For the lowest mass clusters, the mass bias is particularly sensitive to the fit radii and the limits placed on the concentration prior, rendering reliable mass estimates difficult. The scatter in mass estimates between the dark matter-only and the various baryonic runs is less than between different projections of individual clusters, highlighting the importance of triaxiality.
\end{abstract}




\section{Introduction}

In the 1930s, Fritz Zwicky discovered that most of the mass in galaxy clusters is in the form of dark matter \citep{zwicky1937}. After the advent of X-ray telescopes, it was established that X-ray emitting plasma comprises most of the normal luminous mass in clusters. Our current best estimate is that typical clusters are composed of more than 80\% dark matter, about 17\% plasma, and at most a few percent in the form of stars in galaxies \citep[e.g.,][]{allen2011}.

Since their properties provide important tests of our cosmological model and our understanding of structure formation, galaxy clusters are targets of many ongoing and upcoming surveys. Clusters are also testbeds for investigating the large-scale properties and physics of dark matter and the more complex physics of luminous matter \citep{randall2008}.

Estimating cluster masses from survey observables is crucial to using clusters as cosmological probes \citep[for a review of cluster observations in the context of cosmology see][]{allen2011}. The two key steps in this are relative calibration of cluster masses and absolute calibration of cluster masses. The former involves identifying cluster observables from multi-wavelength data that provide a low-scatter proxy for cluster mass. X-ray measurements, for example, show a scatter of ~10 per cent \citep[e.g.,][]{kravtsov2006}. Even if only a small fraction of clusters have associated low-scatter mass-proxy measurements, constraints on cosmology are improved by a factor of a few \citep[e.g.,][]{wu2011}. For absolute mass calibration, weak gravitational lensing measurements are critical. Gravitational lensing, which uses strongly and weakly lensed background galaxies, is well-established as a tool to map and weigh galaxy clusters. Lensing enables masses to be estimated irrespective of whether the matter is luminous or dark, and without assumptions about a cluster's dynamical state. 

 \citet*{navarro1996} used computer simulations of cosmological structure formation to determine that haloes formed from cold dark matter (CDM) are well described by a particular family of density profiles known as the NFW profile. Due to the computational intensity of modeling baryons, many cosmological simulations include only the gravitational forces between particles. Studies of such dark matter-only clusters have shown that weak lensing masses typically underestimate the true mass by $\approx 5-10\%$, with the bias improving for higher mass \citep{beckerkravtsov2011,oguri2011,bahe2012}. Analyses of more recent, high-resolution simulations have shown that the Einasto profile \citep{einasto1965} tends to be a better fit to simulated clusters, particularly in the inner regions. However, given the fact that observers typically exclude the inner, strong lensing region, the Einasto profile only slightly improves the weak lensing mass bias \citep{henson2017}.
 
More recently, simulations of structure formation have either included baryons and their more complex physics, or they have extracted massive haloes from dark matter-only simulations and re-simulated them with baryonic processes. Although dark matter dominates the mass of clusters, hydrodynamic simulations have shown that baryons can have a significant impact on the structure of low-mass clusters and result in fewer high-mass clusters \citep{henson2017,velliscig2014, velliscig2015, bryan2013, martizzi2012,mummery2017, cusworth2014}. Simulations neglecting AGN feedback suffer from over-cooling, since in the absence of efficient heating the central regions of such clusters have overly efficient heat dissipation, resulting in much higher stellar fractions than observed \citep[e.g.,][]{borgani2011}.

Constraints on the nature and properties of dark energy may be derived from a census of the number of clusters as a function of mass and redshift. Often, the results of surveys are interpreted by comparison with cosmological simulations carried out with only dark matter. However, recent work with hydrodynamical simulations has shown that cosmological parameter estimations calculated from either the halo mass function \citep{velliscig2014,cusworth2014,bocquet2016} or the matter power spectrum \citep{semboloni2011,vandaalen2011} are sensitive to the presence of baryons at the per cent level. Therefore, in the era of precision cosmology, baryons must be accounted for. While the presence of baryons is sub-dominant, we consider the impact they have on the determination of cluster mass from gravitational lensing studies. 

In this paper, we assess how baryonic physics modifies the distribution and estimation of cluster mass with respect to clusters composed of only dark matter. We consider how the impact of baryons on cluster mass estimation compares with intrinsic factors such as cluster shape and choices of fit radius and priors on cluster parameters during data analysis. 

In Section \ref{s:wl}, we describe the NFW density profile and general weak lensing formalism. In Section \ref{s:sims}, we describe the cosmo-OWLS \citep{lebrun2014} hydrodynamical simulation suite from which our cluster sample is extracted, as well as our process for producing mock weak lensing observations of these clusters. We then discuss our methods of mass estimation in Section \ref{s:mass_est}. We use a Markov Chain Monte Carlo (MCMC) algorithm to compute the posterior probability distribution for the lensing $M_{200}$ of individual clusters. These distributions then serve as input to an additional MCMC step to calculate the bias and scatter of the mass estimates. In Section \ref{s:results}, we present mass estimations of our cluster sample for two different noise levels and three inner fit radii. We also show the sensitivity of cluster mass estimation on the concentration prior, and additionally the relative uncertainty of parameter estimation due to prescriptions of baryonic physics as compared to projection effects. We discuss the results and future work in Section \ref{s:conc}.

\section{NFW density profile and weak lensing}
\label{s:wl}
\subsection{NFW density profile}
As is common in weak lensing surveys, we obtain mass and concentration estimates using the spherically symmetric NFW profile described in \citet{navarro1996}. This allows a straightforward comparison between the true parameters of the simulated clusters and those derived from lensing observables. The NFW profile has analytic expressions for the convergence and shear. It is given as a function of radius, $r$, as
\begin{equation}
\label{nfwprofile}
\rho_{NFW}(r) = \frac{\delta_c \rho_c}{(r/r_s)(1+r/r_s)^2}
\end{equation}
where $\rho_c$ is the critical density of the universe, and the parameters $\delta_c$ and $r_s$ are the cluster's central over-density and the scale radius, respectively. We may parameterize the profile with the dimensionless concentration parameter, $c$, which satisfies
\begin{equation}
\delta_c = \frac{200}{3} \frac{c^3}{\ln{(1+c)} - c/(1+c)}
\end{equation}
and the radius $r_{200}=r_s \cdot c$, which defines a sphere within which the average density is $200 \cdot \rho_c(z)$. The mass contained within this sphere, $M_{200}$, is then given by
\begin{equation}
M_{200} = M(r<r_{200}) = \frac{800\pi}{3} \rho_c(z)r^3_{200}.
\end{equation}
$M_{500}$ is similarly defined as the mass contained within a sphere in which the average density is $500 \cdot \rho_c(z)$.
In this form, the NFW profile has two parameters. It is possible to reduce this to a one parameter problem by utilizing a power-law mass-concentration ($M$-$c$) relation. However, studies have shown that $M$-$c$ relations are sensitive to baryons, since the inclusion of efficient feedback results in reduces the NFW concentration and the amplitude of the $M$-$c$ relation \citep[e.g.,][]{duffy2010,mummery2017}. We therefore allow the concentration parameter to vary freely in order to capture the changes in concentration resulting from the inclusion of baryonic physics. It may also be possible to include a $M$-$c$ relation prior broad enough to capture the impact of baryons, but we do not attempt this in this study.

\subsection{Weak lensing formalism}
We consider the cluster as a two-dimensional lens, where the surface mass density can be obtained by integrating the three dimensional density along the line-of-sight,
\begin{equation}
\Sigma(\vec{r}) = \int^{\infty}_{-\infty} \rho(\vec{r},z)dz
\end{equation}
where $\vec{r}$ is a position vector in the lens plane.
In the standard weak lensing notation, $\kappa(\vec{r})$ denotes the convergence, i.e. the dimensionless surface mass density of the lens. It is defined as $\kappa(\vec{r}) = \Sigma(\vec{r})/\Sigma_c$, where
\begin{equation}
\Sigma_c = \frac{v_c^2}{4\pi G} \frac{D_s}{D_d D_{ds}}
\end{equation}
is the critical mass density \citep{subramanian1986}. $D_d$, $D_s$, and $D_{ds}$ are the angular diameter distances from the observer to the lens, the observer to the source, and the lens to the source, respectively. $v_c$ is the speed of light and $G$ is the gravitational constant. The convergence is related to the deflection potential, $\psi(\vec{r})$, through the Poisson equation, $\nabla^{2}\psi = 2\kappa$.

The deflection potential is also related to the complex-valued shear, through $\gamma=\gamma_1 + i\gamma_2 = \psi,_{11} - \psi,_{22} + i\psi,_{12}$ where the indices denote the derivatives with respect to the position in the lens plane. Noting that $\kappa$ and $\gamma$ are combinations of second derivatives of the potential, the convergence and shear may easily be related to each other by use of the Fourier transform,
\begin{equation}
\label{fourierrel}
\tilde{\gamma} = \left( \frac{\hat{k}^2_1 - \hat{k}^2_2}{\hat{k}^2_1+\hat{k}^2_2} \tilde{\kappa}, \frac{2\hat{k}_1 \hat{k}_2}{\hat{k}^2_1+\hat{k}^2_2} \tilde{\kappa} \right),
\end{equation}
where $\tilde{\gamma}$ and $\tilde{\kappa}$ respectively denote the shear and convergence in Fourier space, and $\hat{k}_i$ denotes the components of the corresponding wave vector. For the case of the spherical NFW profile, the convergence and shear may be calculated analytically \citep{bartelmann1996}.

In the regime $\kappa,\gamma \ll 1$, the lensing signal is found by measuring the distortion by the deflector of the the source galaxy ellipticities. We use the complex-valued ellipticity $\epsilon$ with modulus $\left| \epsilon \right| =  (1-q)/(1+q)$, where $q$ is the axis ratio. The locally linearized relationship between the intrinsic source ellipticity, $\epsilon^s$, and the observed image ellipticity, $\epsilon$, is 

\begin{equation}
\label{e:imellip}
\epsilon = \frac{\epsilon^s + g}{1+g^*\epsilon^s},
\end{equation}
where $g=\gamma/(1-\kappa)$ is the reduced shear, and $g^*$ indicates the complex conjugate of $g$. Since the galaxies are randomly distributed, $\left<\epsilon_s\right>=0,$ hence the locally averaged image ellipticities are related to the reduced shear through
\begin{equation}
\left< \epsilon \right> \approx \left< g \right> + \left< \epsilon^s \right> = \left< g \right>.
\end{equation}

\section{Simulations}
\label{s:sims}

\subsection{Cosmo-OWLS}

\begin{figure}
\includegraphics[width=\linewidth]{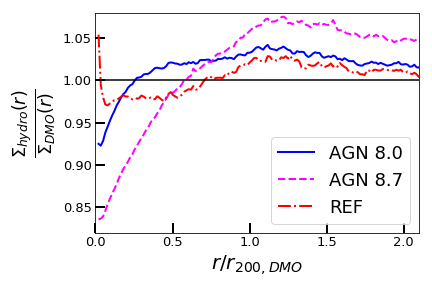}\par
\caption{This figure shows the mean ratio of the projected density profiles of all baryonic clusters to their corresponding DMO clusters as a function of $r/r_{200}$, where the $r_{200}$ are from the 3D spherical over-density mass definition.}
\label{profile_compare}
\end{figure}

\begin{figure*}
\begin{multicols}{2}
\includegraphics[width=\linewidth]{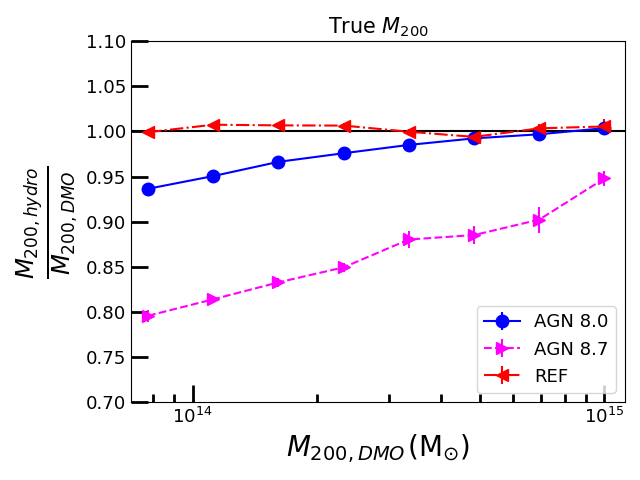}\par
\includegraphics[width=\linewidth]{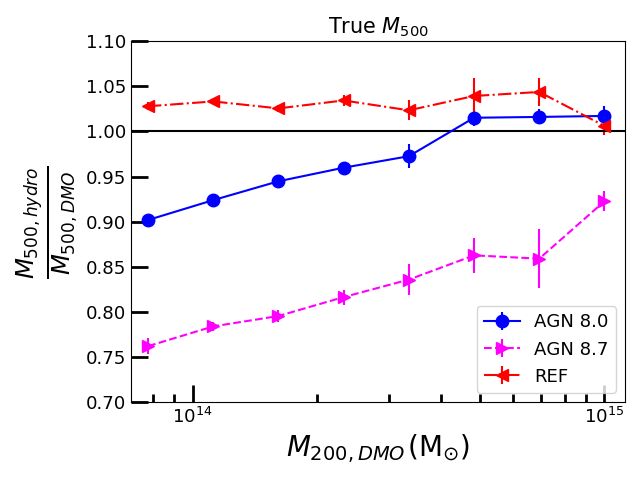}\par
\end{multicols}
\caption{The panel on the left shows the ratio of $M_{200}$ of the baryonic clusters to their DMO counterparts, where the $M_{200}$ are calculated using the 3D spherical over-density mass definition. The panel on the right shows the similarly for $M_{500}$. Both are as a function of the $M_{200}$ of the corresponding DMO cluster to maintain the same bins.}
\label{M_compare}
\end{figure*}

Cosmo-OWLS, described in \citet{lebrun2014}, is a suite of large-scale hydrodynamical simulations consisting of periodic boxes of 400 $h^{-1}$ comoving Mpc on a side. The simulations use initial conditions based on maximum likelihood parameter values derived from either \textit{Planck} \citep{planck2014} or \textit{WMAP7} data \citep{wmap7}. In this work, we use runs produced with the \textit{WMAP7} cosmology, where $\{\Omega_m,\Omega_b,\Omega_{\Lambda},\sigma_8, n_s, h\} = \{0.272,0.0455,0.728,0.81,0.967,0.704\}$. Each run contains $2\times1024^3$ particles with masses $\approx 3.75 \times 10^9 h^{-1} \mathrm{M_{\odot}}$ and $\approx 7.54 \times 10^8 h^{-1} \mathrm{M_{\odot}}$ for dark matter particles and baryons, respectively. The prescription of \citet{eisenstein1999} was used to compute the transfer function, and the software \textsc{N-GenIC} was used to generate the initial conditions.

The simulations were run using a version of the Lagrangian TreePM-SPH code \textsc{gadget3} \citep[last described in][]{gadget2}. The haloes were identified after running an on-the-fly friends-of-friends (FOF) algorithm with a linking length of 0.2 times the mean interparticle separation. The spherical over-density masses and radii for each of the FOF haloes were calculated using spheres centered on the halo's most-bound particle.

We use four of the six runs produced by \citet{lebrun2014}, all of which used identical initial conditions. The models we use are:
\begin{itemize}
	\item DMO: a dark matter-only run that simulates only the gravitational interaction between particles.
	\item REF:  in addition to gravity, this run also implements a UV/X-ray photoionizing background \citep{haardt2001}, element-by-element radiative cooling \citep{wiersma2009}, star formation \citep{schaye2008}, stellar evolution and chemical enrichment \citep{wiersma2009b}, and kinetic wind from supernova feedback \citep{dallavecchia2008}.
	\item AGN 8.0 and 8.7: these runs implement all of the processes included in REF with the addition of the growth of supermassive black holes (BH) and AGN feedback \citep{springel2005, boothschaye2009}. During the simulation, FOF haloes with at least 100 dark matter particles were seeded with a BH with an initial mass of 0.001 times the initial gas particle mass. The BHs were allowed to grow either through mergers with other BH particles or by scaled Eddington-limited Bondi-Hoyle-Lyttleton accretion as described in \citet{boothschaye2009}. Initially, the BHs store this additional energy until they gain enough mass so that they are able to increase the temperature of neighbouring gas particles by a pre-defined temperature, $\Delta T_{heat}$. This temperature is set to $\Delta T_{heat} = 10^{8.0}$ K and $\Delta T_{heat} = 10^{8.7}$ K for the AGN 8.0 and 8.7 models, respectively. There is additionally an AGN 8.5 model with $\Delta T_{heat} = 10^{8.5}$ K, but we do not analyze this in this work.
\end{itemize}

In comparison with \citet{henson2017}, who studied one implementation of baryonic physics in addition to the DMO run, we consider three such implementations. This allows us to consider a wide range of possibilities for the impact of AGN feedback and other physical processes.

\citet{lebrun2014} found that several observables, such as various global hot gas properties and the density profiles of the ICM, are bracketed by AGN 8.0 and AGN 8.5 when assuming the \textit{WMAP7} cosmology that is adopted here. \footnote{Note that in the \textit{Planck} cosmology, AGN 8.0 best reproduces the observables.} The AGN 8.7 run is a somewhat extreme model, yielding significantly lower gas mass fractions than observed. Thus, AGN 8.0 and AGN 8.7 encompass the possible cluster observables. REF omits AGN feedback entirely, resulting in clusters with extremely high central densities from cooling. AGN 8.7 and REF are included in this study for comparison as different extremes of AGN feedback. The cosmo-OWLS suite has been employed to study, the alignments of baryonic components with dark matter haloes \citep{velliscig2015}, the effects of baryon physics on large-scale structure \citep{mummery2017}, the implications of the Planck CMB best-fit parameters on the thermal Sunyaev-Zel'dovich (tSZ) effect power spectrum \citep{mccarthy2014}, and the tSZ-lensing cross-correlation \citep{hojjati2015}.

\subsection{Cluster selection and matching}

The haloes included in our sample were selected by first extracting all the clusters above $10^{14} \mathrm{M_{\odot}}$ at $z=0.25$ from the DMO run in boxes 30 Mpc on a side. The size of the extracted boxes was selected to include correlated large-scale structure (e.g., filaments connected to the clusters). Both \citet{bahe2012} and \citet{beckerkravtsov2011} found little variation in weak lensing mass estimates for LOS integration lengths from 10 $h^{-1}$ Mpc to 50 $h^{-1}$ Mpc for the mass range we consider here. Much longer integration lengths to include uncorrelated large-scale structure would require the use of ray-tracing algorithms which is beyond the scope of this work.

The haloes were matched between the runs by finding clusters located within the distance of five times the candidate cluster's $r_{500}$ in DMO and were within a factor a third of its $M_{500}$. This simple selection criterion allowed for 95 per cent of the clusters to be uniquely matched between the four runs. The remaining 5 per cent mostly occurred when two clusters from a baryon run satisfied the matching criteria for a given DMO cluster. In these cases, we checked the positions and masses of the candidate clusters and found that one was always much closer to the corresponding DMO cluster's location and mass than the other. We therefore included these clusters in our sample. The remaining 5 cases were low-mass clusters at the edge of the simulation box which failed to match with any candidate clusters. We discarded these from our sample. The total number of clusters in our sample for each of the simulations is 1,157.

Figure \ref{profile_compare} shows the relative density profiles of the different baryon runs with respect to DMO, and Figure \ref{M_compare} shows the differences in cluster $M_{200}$ and $M_{500}$ as a function of the $M_{200}$ of the matched DMO clusters. Clusters from AGN 8.0 and 8.7 tend to have lower inner densities, with some of the matter being pushed outward to the cluster periphery. This also results in a smaller $M_{200}$, with the effect more pronounced when the AGN feedback is stronger. This impact declines for higher mass, with little mass difference between high-mass AGN 8.0 and DMO clusters. Clusters in REF are dense toward the center, though beyond $r/r_{200} \approx 0.1$ matter tends to be pushed out, which is due to SN feedback as determined by \citet{lebrun2014}. The $M_{500}$ of the AGN models are more suppressed with respect to DMO, though the overall trend is similar to the $M_{200}$. This is due to the fact that $M_{500}$ corresponds to a smaller radius than $M_{200}$, so this mass definition does not capture as much of the matter cast into the periphery from AGN feedback. The REF clusters tend to have slightly higher $M_{500}$ compared to their DMO counterparts. This is because the REF clusters tend to have higher central densities, and because some of the regions in which REF clusters are less dense than DMO clusters are outside of the $r_{500}$.

\subsection{Catalog generation}

Each of the extracted clusters were projected along three orthogonal axes to enhance our sample size, as well as to provide information about the scatter in mass estimations between projections. We generate convergence maps by dividing each surface mass density map by $\Sigma_c$ calculated for a cluster redshift of $z_l=0.25$ and a source redshift of $z_s=1$ in the \textit{WMAP7} cosmology. We then produced shear maps by using the relationship in Fourier space described in equation \ref{fourierrel}.

For each projection, a set of randomly placed background galaxies at redshift $z_s=1$ were generated. Following \citet{geiger1998}, the ellipticity components of each galaxy were randomly drawn from the distribution,
\begin{equation}
\label{e:intdist}
p^s_{\epsilon}(\epsilon^s) = \frac{\exp(-|\epsilon^s|^2/\sigma^2_\epsilon)}{\pi\sigma^2_\epsilon(1-\exp(-1/\sigma_\epsilon^2))},
\end{equation}
where $\sigma_\epsilon$ is the dispersion of the intrinsic ellipticity distribution, which we set to $\sigma_\epsilon = 0.05$ when studying algorithm performance at low noise levels and $\sigma_\epsilon=0.25$ for more realistic noise.

We use an unlensed galaxy number density of $n_0 = 30$ gal/arcmin$^2$, the approximate expected number density of the LSST in the absence of lensing magnification. The total number of galaxies is a realization of a Poisson distribution with mean $N=n_0A$, where $A$ is the area of the sky. Following \citet{schneider2000}, the galaxy number density must also be adjusted due to the effects of lensing magnification. While Equation \ref{e:imellip} describes the distortion of the source shapes, we must also numerically account for the magnification of space around the sources as well as the change in the lensed galaxies' brightness. The lensed number density satisfies
\begin{equation}
n(\vec{r}) = n_0 \mu^{-0.5},
\end{equation}
where $\vec{r}$ is again a position vector in the lens plane and $\mu$ is the magnification at that position. $\mu$ is calculated for each source galaxy's position from the lens shear and convergence. A uniform variate $\eta \in [0,1]$ is drawn, and the galaxy is only included in the catalog if $[\mu]^{-1/2} \geq \eta$; otherwise, it is discarded.

\section{Mass estimation}
\label{s:mass_est}
\subsection{Likelihood function}

To calculate the best-fit values and the confidence intervals for the cluster parameters, we make use of the likelihood and log-likelihood functions. The likelihood function is defined as the probability that a set of parameters will result in the given observables. The likelihood and log-likelihood functions are given by
\begin{equation}
\label{e:likeeqs}
\begin{split}
\mathcal{L} &= \prod_{i=1}^{N_{gal}} p_{\epsilon}(\epsilon_i \vert g(\vec{x}_i))
\\
\ell &= \sum_{i=1}^{N_{gal}} \ln p_{\epsilon}(\epsilon_i \vert g(\vec{x}_i)),
\end{split}
\end{equation}
where $\epsilon_i$ and $\vec{x}_i$ are respectively the measured image ellipticity and image position in the lens plane, and $N_{gal}$ is the total number of source galaxies for which the image ellipticities have been measured. 
From \citet{geiger1998}, the probability distribution of the image ellipticities can then be written as
\begin{equation}
\label{e:imdist}
p_{\epsilon}(\epsilon \vert g) = p^s_{\epsilon}(\epsilon^s(\epsilon \vert g)) \left| \frac{\partial^2\epsilon^s}{\partial^2\epsilon}\right| = p^s_{\epsilon}(\epsilon^s(\epsilon \vert g)) \frac{(|g|^2-1)^2}{\left| \epsilon g^*-1\right| ^4} .
\end{equation}

Substituting equations \ref{e:intdist} and \ref{e:imdist} into equation \ref{e:likeeqs}, the likelihood function becomes
\begin{equation}
\mathcal{L} = \prod_{i=1}^{N_{gal}} \frac{\exp\left( -\vert \epsilon^s_i \vert^2/\sigma_{\epsilon}^2 \right)}{\pi\sigma_{\epsilon}^2(1-\exp\left(-1/\sigma_{\epsilon}^2\right))} \frac{(|g_i|^2-1)^2}{\left| \epsilon_i g_i^*-1\right| ^4},
\end{equation}
where $\epsilon^s_i$ is included by inverting equation \ref{e:imellip} for each of the galaxies,
\begin{equation}
\epsilon_i^s = \frac{\epsilon_i - g_i}{1-g_i^*\epsilon_i}.
\end{equation}
The log-likelihood function is then
\begin{equation}
\label{loglike}
\begin{split}
\ell = \sum_{i=1}^{N_{gal}} \ln\left[ (\vert g_i \vert^2 - 1)^2\right] & -\frac{\vert \epsilon^s_i \vert^2}{\sigma^2_\epsilon} -\ln \vert \epsilon_i g_i^* - 1 \vert ^4,
\end{split}
\end{equation}
where the constant additive term has been dropped. Note that the $\epsilon^s_i$ in the likelihood is what the $i$th galaxy's intrinsic ellipticity would be for a given set of model parameters and the observed image ellipticity, since the true intrinsic ellipticity is unknown.

\subsection{MCMC}
\label{s:mcmc}

Our method of cluster parameter estimation involves the use of Markov Chain Monte Carlo (MCMC) which is based on Bayesian statistics. Bayes' theorem states that given the data, $\mathrm{\mathbf{D}}$, the probability distribution of a set of parameters, $\Theta$, is
\begin{equation}
\label{bayestheorem}
p(\Theta \vert \mathrm{\mathbf{D}}) = \frac{p(\mathrm{\mathbf{D}}\vert \Theta) p(\Theta)}{p(\mathrm{\mathbf{D}})},
\end{equation}
where $p(\Theta \vert \mathrm{\mathbf{D}})$ is the posterior probability distribution, $p(\mathrm{\mathbf{D}}\vert \Theta)$ is the likelihood, $p(\Theta)$ is the prior, and $p(\mathrm{\mathbf{D}})$ is the evidence.

MCMC methods involve a guided random walk through the posterior probability distribution, returning a representative set of sample points. The density of sample points in a region of the posterior is directly proportional to the probability that the region contains the true parameter value. This sample subset approximates the full posterior distribution, which can then be marginalized over to obtain the mean most likely parameter values without making \textit{a priori} assumptions about the form of the posterior.

In general, the MCMC sampler walks through parameter space, with steps governed by some transfer function which depends on the particular method employed. We use \textsc{emcee}, the Python implementation of the affine-invariant MCMC ensemble sampler \citep{goodman2010,emcee}. In short, rather than producing a single MCMC chain, this method utilizes multiple `walkers', each exploring parameter space from different starting positions. Candidate samples are selected for each walker using the `stretch move', wherein a point is chosen along the vector between the current position of the walker and of another randomly selected walker. One of the benefits of this method is that the autocorrelation time, an important test of convergence, tends to be much shorter than for the generic Metropolis-Hastings method. Additionally, affine-invariant samplers perform equally well when there is covariance between parameters. A more detailed explanation of \textsc{emcee} and of affine-invariant ensemble samplers can be found in \citet{emcee} and \citet{goodman2010}, respectively.

For our particular problem, we wish to obtain a representative set of samples from $p(M_{200},c)$, the posterior probability distribution for the mass and concentration. For each individual cluster, the dataset on which we do the analysis stays constant, so we may neglect the evidence term in \ref{bayestheorem}. We impose flat priors for both $M_{200}$ and $c$ such that
\begin{equation}
\begin{split}
c &\in \mathcal{U}(1,30) \\
M_{200} &\in \mathcal{U}(\sim 1.45 \times 10^{11} \mathrm{M_{\odot}}, \sim 1.81 \times 10^{16} \mathrm{M_{\odot}}).
\end{split}
\end{equation}
The inner and outer limits on $M_{200}$ correspond to $r_{200} = 0.1$ Mpc and 3 Mpc, respectively. We use the log-likelihood function from equation \ref{loglike}. Using the MCMC sampler previously described above, we start 20 walkers centered in a tight Gaussian ball around the best-fit maximum likelihood mass and concentration calculated using the previously produced mock WL catalogs. Each walker has a 200 step burn-in period, and the following 500 samples are combined across all walkers for a total of 10000 samples from the posterior distribution for each cluster projection. We then make use of these MCMC chains to calculate the bias and scatter in $M_{200}$ as described in the next section. To ensure that parameter space has been properly explored by the MCMC algorithm, we perform a test by selecting ten clusters at random from our sample, each with three projections for a sample size of thirty. We place the twenty walkers at random around parameter space within the allowed range of the priors and verify that the mean change in the estimated masses is less than 1 per cent.

\subsection{Bias and scatter calculation}
\label{s:bias}

We again make use of MCMC methods to calculate the various statistics of mass estimates for each mass bin by utilizing the MCMC chains produced for each cluster. For a given set of data, $\vec{g}_i$, corresponding to the $i$th cluster in a mass bin, the probability distribution of the resulting mass bias is
\begin{equation}
\label{bias_like_single}
p_i\left(\mu, \sigma | \vec{g}_i \right) \propto \left[\int dM_{lens}p(\vec{g}_i|M_{lens})p(M_{lens}|\mu, \sigma)\right] p(\mu, \sigma),
\end{equation}
where $\mu$ and $\sigma$ are measures of bias and scatter in the estimated mass (denoted here as $M_{lens}$). Applying Bayes' theorem to $p(\vec{g}_i|M_{lens})$, we have
\begin{equation}
p_i(\vec{g}_i|M_{lens}) \propto \frac{p(M_{lens}|\vec{g}_i)}{p(M_{lens})}.
\end{equation}
Since the mass prior, $p(M_{lens})$, is constant in the range of allowed $M_{lens}$, neglecting it yields no difference in equation \ref{bias_like_single}. The distribution $p(M_{lens}|\vec{g}_i)$ is the posterior of the estimated mass from which we already sampled following Section \ref{s:mcmc}. We expect $p(M_{lens}|\mu, \sigma)$ to take the form of a log-normal distribution \citep[see discussion in][]{stanek2010} and implementation in \citet{schrabback2016}). The probability distribution of $M_{lens}$ is then
\begin{equation}
p\left(M_{lens} | \mu,\sigma\right) = \frac{1}{M_{lens}\sqrt{2\pi\sigma^2}} \exp{\left(-\frac{(\ln{M_{lens}} - \ln\mu M_{true})^2}{2\sigma^2}\right)}.
\end{equation}
This equation allows us to interpret $\ln\mu$ and $\sigma^2$ as respectively the mean and variance of the normally distributed $\ln\left(M_{lens}/M_{true}\right)$. We may translate these variables into the various statistical quantities we wish for the corresponding log-normally distributed variable $M_{lens}/M_{true}$. Its median, mean, and variance are respectively
\begin{equation}
\begin{split}
med &= \exp(\ln\mu) = \mu \\
mean &=  \exp\left( \mu + \sigma^2/2 \right)\\
var &= \left[ \exp\left( \sigma^2 \right) -1 \right] \exp \left( 2\mu + \sigma^2 \right).
\end{split}
\end{equation}
The remaining term in equation \ref{bias_like_single}, $p(\mu, \sigma)$, represents the prior on $\mu$ and $\sigma$ which is both independent of the mass samples and is the same for each cluster.

Again using Bayes' theorem, the probability that all the clusters in a bin have a particular bias and scatter satisfies
\begin{equation}
p\left( \mu, \sigma | \vec{g} \right) \propto p\left( \vec{g} | \mu, \sigma\right) p\left(\mu, \sigma) = \mathcal{L}(\mu, \sigma\right) p\left(\mu, \sigma\right),
\end{equation}
where $\mathcal{L}(\mu, \sigma)$ is the likelihood function for $\mu$ and $\sigma$ given the MCMC mass samples for all the clusters in the bin. To obtain $\mathcal{L}(\mu, \sigma)$, we multiply the probabilities $p_i \left( \vec{g}_i| \mu, \sigma \right)$ for all of the clusters in the mass bin. Each of the $p_i \left( \vec{g}_i| \mu, \sigma \right)$ may be identified as the term in the square brackets of equation \ref{bias_like_single}. Replacing the integral with a summation and substituting in the mass samples, the likelihood function for $\mu$ and $\sigma$ is therefore
\begin{equation}
\label{bias_likelihood}
\begin{split}
&\mathcal{L}(\mu, \sigma) = \prod_{i=1}^{N_c} p_i \left( \mu, \sigma | \vec{g}_i\right) \\
 &= \prod_{i=1}^{N_c} \left\{ \frac{1}{N_s}  \sum_{j=1}^{N_s}  \left[ \frac{1}{M_{lens,j,i}\sqrt{2\pi\sigma^2}} \times \right. \right. \\
 &\left. \left. \exp \left( -\frac{(\ln M_{lens,j,i} - \ln \mu M_{true,i})^2}{2\sigma^2} \right) \right] \right\},
\end{split}
\end{equation}
where $N_c$ is the number of clusters in a given bin, $N_s$ is the number of samples, $M_{lens,j,i}$ is the $j$th mass sample of the $i$th cluster's MCMC chain, and $M_{true,i}$ is the cluster's true spherical over-density mass. In practice, we use the natural logarithm of equation \ref{bias_likelihood}. We assume flat priors on $\ln \mu$ and $\ln \sigma$ such that 
\begin{equation}
\begin{split}
\ln \mu &\in \mathcal{U}(-1, 1) \\
\ln \sigma &\in \mathcal{U}(\ln 0.05, \ln 10).
\end{split}
\end{equation}

There are many differences between our approach and that of \citet{henson2017}, who also investigated the impact of baryonic matter on weak lensing mass estimates. Apart from considering a wider range of prescriptions of baryonic physics, we directly fit to the ellipticities of background galaxies in synthetic lensing catalogues following the procedure of \citet{bahe2012} rather than fitting to ideal azimuthally averaged shear profiles. This then includes the impact of finite sampling as well as shape noise. This also allows us to estimate the confidence contours on parameters in the mass model which we do using an MCMC approach.

\begin{figure*}
\includegraphics[scale=1.]{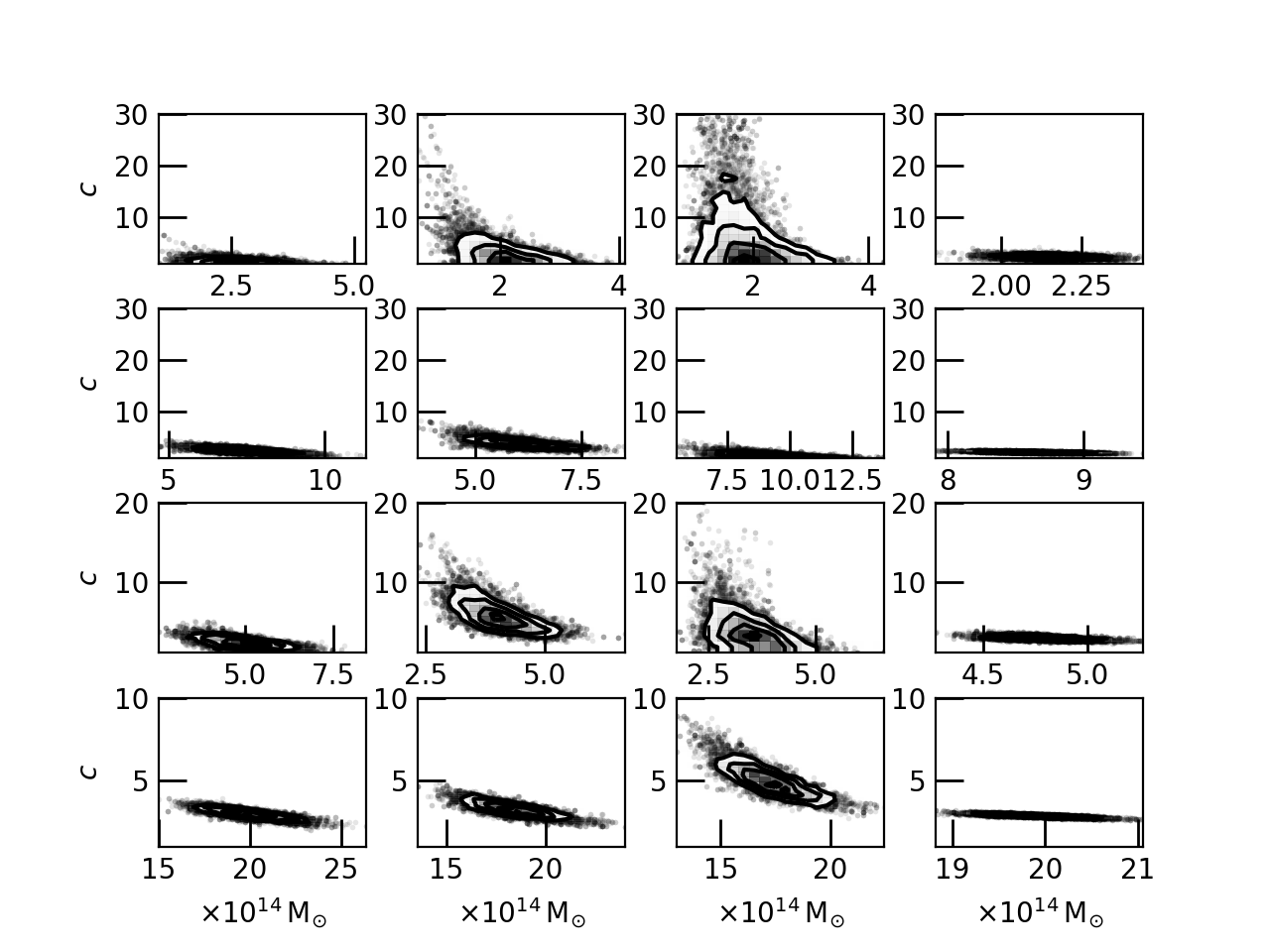}
\caption{This figure shows the distribution of fit mass and concentration samples for 4 example clusters across the mass range.  The left three panels on each row correspond to $\sigma_{\epsilon} = 0.25$ and $r_{in} = 0.1, 0.25,$ and $0.5$ Mpc for the leftmost panels, the centre-left panels, and the centre-right panels respectively. The rightmost panel corresponds to $r_{in}=0.5$ Mpc and $\sigma_{\epsilon} = 0.05$. We do not include the other low-noise distributions as they are all similar. The mass range covered is the same along each row. The concentration range is the same except for the rightmost panels which are zoomed in. This figure was produced using the \textsc{corner} python module \citep{corner}.}
\label{cluster_example_chains}
\end{figure*}

\section{Results}
\label{s:results}

We calculate the bias and scatter in the MCMC fits to our cluster sample using a variety of fit radii and two different noise levels. We set the intrinsic source ellipticity to $\sigma_{\epsilon} = 0.05$ for low-noise fits and $\sigma_{\epsilon} = 0.25$ for a more realistic intrinsic source ellipticity dispersion. To test sensitivity to the inner fit radius, we perform three separate fits for inner fit radii set to $r_{in}=0.1,$ $0.25,$ and $0.5$ Mpc with an outer fit radius of $r_{out}= 3$ Mpc. This is motivated by, for example, \citet{gao2008} who found that the fit radius has an impact on the NFW concentration when fitting directly to spherically averaged density profiles of dark matter haloes from cosmological simulations. It is also motivated by \citet{bahe2012}, who demonstrated that the extent of the data field over which lensing data analysis is performed impacts on the estimated NFW parameters using dark matter-only cosmological simulations. Since baryonic physics impacts the concentration of mass relative to dark matter-only simulations, we examine how varying the inner fit radius changes NFW parameter estimates.

We begin by producing MCMC chains of mass and concentration for each of the cluster projections as described in Section \ref{s:mcmc}, allowing mass and concentration to vary freely. Figure \ref{cluster_example_chains} shows confidence contours for several example clusters across the entire mass range with varying inner fit radii and intrinsic ellipticity dispersions (e.g., see $x$-axis of Figure \ref{M_compare} for lowest mass cluster in each bin). At higher masses, the MCMC fits are typically able to constrain $M_{200}$ and $c$ at realistic noise levels regardless of the selection of inner fit radius. When $\sigma_{\epsilon}$ is lowered to 0.05, the parameter constraints narrow as one would expect. The distributions peak at different values since a different population of background galaxies was used for each cluster projection. At lower masses, the parameter constraints significantly worsen, as illustrated in the top rows of Figure \ref{cluster_example_chains}. In these examples, the confidence intervals of the parameters remain open for every tested value of $r_{in}$ when $\sigma_{\epsilon}=0.25$. For these three runs, the MCMC algorithm typically over-samples regions of unrealistically high concentration. These samples often correspond to lower $M_{200}$ values, since for a given signal, higher concentration leads to lower mass estimates due to the degeneracy between mass and concentration. The parameters become well-constrained with low-noise, as we expect given the significantly higher signal-to-noise.

\begin{figure*}
\begin{multicols}{2}
\includegraphics[width=\linewidth]{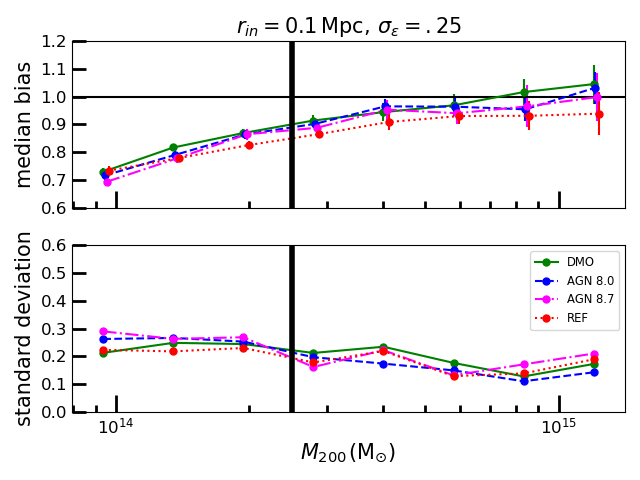}\par
\includegraphics[width=\linewidth]{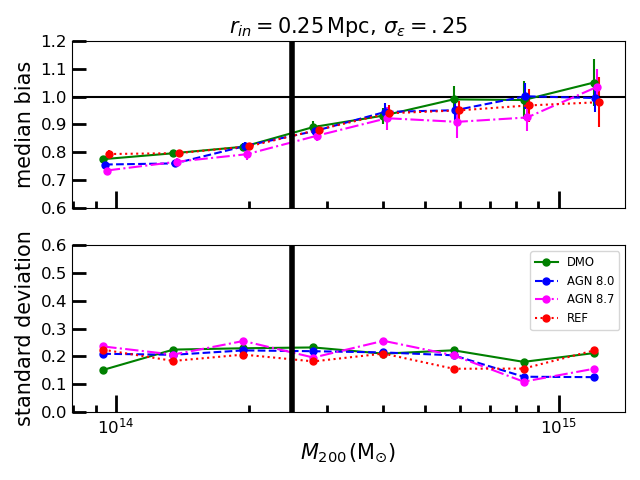}\par
\end{multicols}
\begin{multicols}{2}
\includegraphics[width=\linewidth]{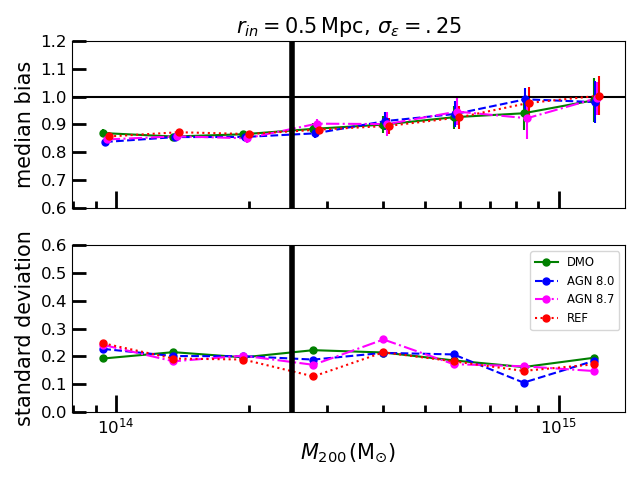}\par
\includegraphics[width=\linewidth]{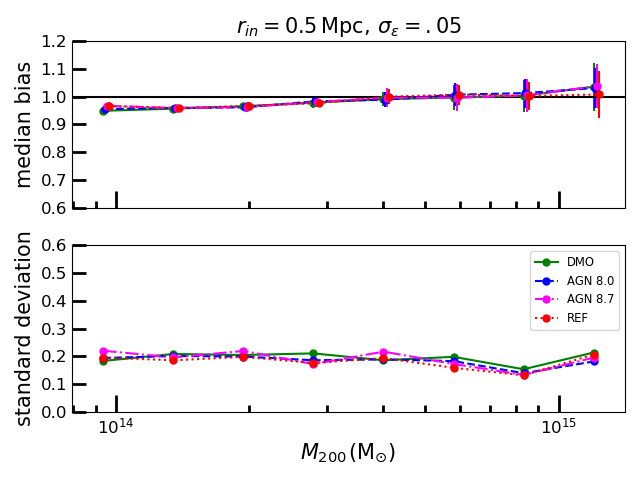}\par
\end{multicols}
\caption{The top left panel of this figure shows the median bias with respect to the true mass and the standard deviation of the estimated masses as a function of true mass using an inner fit radius of $r_{in}=0.1$ Mpc and an intrinsic ellipticity dispersion of $\sigma_{\epsilon}=0.25$. The top right and bottom left panels show the same quantities using an inner fit radius of $r_{in}=0.25$ Mpc and $r_{in}=0.5$ Mpc, respectively. The lower right panel shows the low-noise analog.}
\label{bias_mcmc}
\end{figure*}

We then randomly extract a subset of 2000 samples from each of the individual cluster's MCMC chains to serve as input for the bias and scatter calculations on each mass bin as described in Section \ref{s:bias}, the results of which are shown in Figure \ref{bias_mcmc}. Selecting 2000 samples instead of the full 10000 from each cluster does result in a significant difference in the calculated bias. The top left panel of Figure \ref{bias_mcmc} shows the median bias and standard deviation of the log-normally distributed mass estimates with respect to true mass using an inner fit radius of $r_{in} = 0.1$ Mpc and an intrinsic galaxy ellipticity dispersion of $\sigma_{\epsilon} = 0.25$. For the same galaxy ellipticity dispersion, the top right (bottom left) panel shows the same quantities using an inner fit radius of $r_{in} = 0.25$ Mpc (0.5 Mpc), respectively.  Given the poor constraints on mass and concentration of individual clusters in the lowest mass bins for data of this quality, the first three bins in these subplots can be disregarded. Note that the bias does not significantly vary for different inner fit radii for clusters beyond the lowest mass bins (to the righthand side of the black vertical line in the subplots). The bottom right panel of Figure \ref{bias_mcmc} shows the low-noise analog for an inner fit radius of 0.5 Mpc. Our fits underestimate the $M_{200}$ by $\approx 5\%$ for low masses, consistent with previous literature  \citep[e.g.,][]{bahe2012}. The bias trends slightly upward, closer to 0 per cent for higher masses, though the error bars are larger since the higher mass bins contain fewer clusters. The bias seen in Figure \ref{bias_mcmc} can be explained by the short-comings of the NFW profile in describing cluster profiles, since realistic clusters (such as those considered here from simulations) are often non-spherical, contain substructure, and are embedded in correlated and uncorrelated large-scale structure not captured by the NFW profile \citep[for the factors that cause deviations in estimates of mass and concentration, see][]{bahe2012}.  The low-noise fits for the other fit radii we considered are not shown as they look essentially identical to the lower right panel of Figure \ref{bias_mcmc}.

\begin{figure*}
\begin{multicols}{2}
\includegraphics[width=\linewidth]{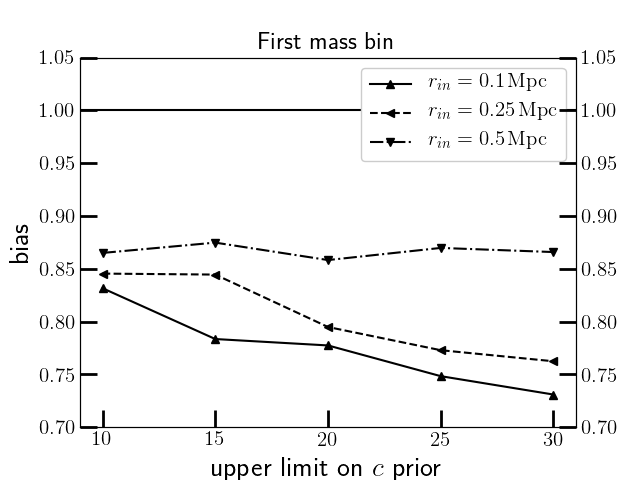}\par
\includegraphics[width=\linewidth]{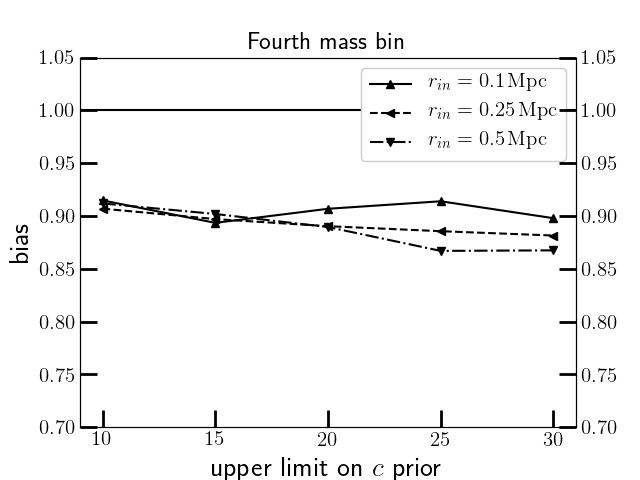}\par
\end{multicols}
\caption{This figure shows the bias sensitivity to the limits placed on the concentration prior for the first (left panel) and the fourth (right panel) lowest mass bins for all inner fit radii considered. By the fourth lowest mass bin, dependence on the concentration prior and inner fit radius disappears.}
\label{ccap_bias}
\end{figure*}

Considering the fourth lowest mass bin, where individual cluster masses are well-constrained, the right panel of Figure \ref{ccap_bias} illustrates that changing the upper bound of the prior on concentration from 30 to 10 changes the bias by only a few per cent. For lower mass bins, where the concentrations of individual clusters are poorly constrained, the limits on the concentration prior and the selection of the inner fit radius significantly impact the resulting bias, shown for the first bin in the left panel of Figure \ref{ccap_bias}. Lowering the upper limit on the concentration prior may artificially improve the mass estimates, as regions of high concentration (and correspondingly low mass) are excluded from the allowed parameter space. Therefore, we do not consider mass estimates below the vertical black line to be reliable.

\begin{figure*}
\begin{multicols}{2}
\includegraphics[width=\linewidth]{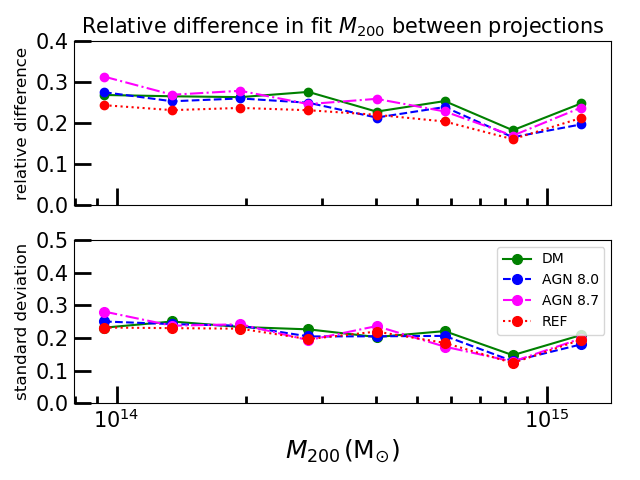}\par
\includegraphics[width=\linewidth]{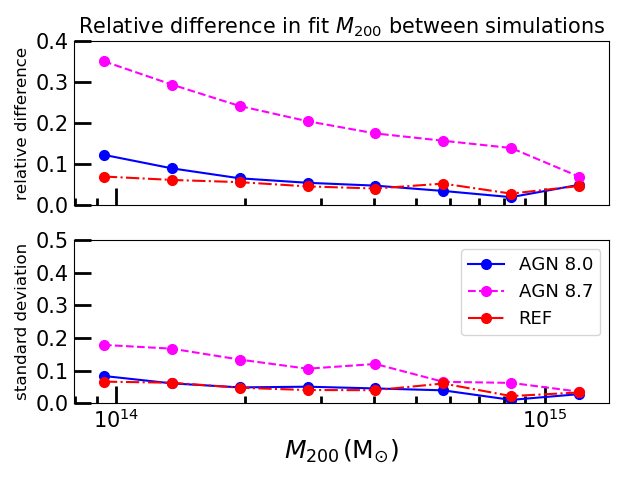}\par
\end{multicols}
\caption{The top left panel of this figure shows the magnitude of the average relative difference in mass estimates between projections for each cluster calculated pairwise. The bottom left panel shows the scatter. The top right panel shows the magnitude of the average relative difference in mass estimates between the matched DMO clusters and those including baryonic physics, and the bottom right panel shows the scatter.}
\label{proj_vs_sim}
\end{figure*}

We then perform fits using the same catalog of galaxies for each projection of each cluster in each simulation run. The left panel of Figure \ref{proj_vs_sim} shows the magnitude of the average relative difference in mass estimates between projections calculated pairwise for each cluster. The right panel shows the relative difference in mass estimates between the clusters containing only dark matter and their corresponding baryonic clusters. These results indicate that scatter from differences in cluster projections (and hence cluster shape) dominate over the scatter due to the presence of baryons: we return to this in the conclusions.

\section{Conclusions}
\label{s:conc}

We have studied the sensitivity of the masses of clusters derived from weak lensing measurements to the presence of baryonic physics in comparison with the presence of dark matter only. Our conclusions are:

\begin{enumerate}
	\item There is no significant difference in WL mass bias and scatter between dark matter-only and galaxy clusters with baryons included, for clusters of mass $M_{200} \geq 10^{14} M_{\odot}$ . This specifically applies to parameterised mass models of the NFW form, obtained while allowing the concentration parameter to vary freely, since M-c relations are sensitive to the presence of baryons.
	\item Lower mass clusters are especially sensitive to the value of the inner fit radius and limits on the concentration prior. Mass and concentration parameters are poorly constrained for individual clusters below $M_{200} \approx 3\times10^{14} M_{\odot}$. Stacking or use of an M-c relation is necessary to obtain parameter estimates for these low-mass clusters.
	\item We confirm the -5 to -10 per cent bias and scatter in NFW mass estimates found in previous work  \citep{beckerkravtsov2011,oguri2011,bahe2012} and show that this holds true for simulated clusters including baryonic processes. For higher mass clusters, we find that this bias improves somewhat.
	\item We confirm that the masses of clusters taken directly from the simulations differ between DMO and the baryonic runs that include AGN feedback, with the latter on average having lower masses. This is consistent with \citet{velliscig2014} and \citet{cusworth2014}. We also demonstrate similar differences between cluster masses derived from the synthetic weak lensing measurements. The difference between the dark matter-only and AGN 8.0 clusters is of order 10 per cent for the lowest mass clusters, tending to zero per cent difference for the highest mass. This is likely a reflection of the greater efficiency of AGN feedback for the lower mass clusters.	 As discussed by e.g. \citet{cusworth2014} this is a very important consideration when cluster mass functions are used as cosmological probes. This will be the topic of a future study. 
	\item We find that differences in cluster projections, a result of triaxial, non-spherical, clusters, tend to dominate over the impact of baryons on WL mass estimation for individual massive clusters. However, there are two important facts to note here. Firstly, while scatter from cluster shape is larger, stacking many clusters together will effectively average out their individual shapes, whereas baryonic processes (specifically AGN feedback), impact all clusters in more or less the same way. Therefore stacking clusters will not smooth out the impact of baryons. Secondly, we emphasize that although the mass estimation bias is unchanged between simulation runs of DMO and various baryonic models, the overall masses of the haloes differ between the simulation runs. This is important for constraints on cosmology from the cluster mass function \citep[e.g.,][]{cusworth2014} and will be the topic of future work.

\end{enumerate}

\section*{Acknowledgements}

BEL and LJK acknowledge support for this work by National Aeronautics and Space Administration Grant No. NNX16AF53G. BEL also acknowledges support from the Texas Space Grant Consortium's Graduate Fellowship program.

AMCLB was supported by the European Research Council under the European Union's Seventh Framework Programme (FP7/2007-2013) / ERC grant agreement number 340519.

DA recognizes support by the Kavli Institute for Cosmological Physics at the University of Chicago through grant NSF PHY-1125897 and an endowment from the Kavli Foundation and its founder Fred Kavli.

BEL thanks Ana Nuñez and Christopher Middleton for proofreading drafts of this paper. We also thank Camille Avestruz for helpful comments on the manuscript.

\bibliographystyle{mnras}
\bibliography{baryonpaper}

\bsp	
\label{lastpage}
\end{document}